# Fast optical modulation of the fluorescence from a single NV centre

Michael Geiselmann [1, &], Renaud Marty [1, &], F. Javier García de Abajo [1,2] *, Romain Quidant [1, 2] *

[1] ICFO - Institut de Ciencies Fotoniques, Mediterranean Technology Park, 08860 Castelldefels (Barcelona), Spain

[2] ICREA - Institució Catalana de Recerca i Estudis Avançats, Barcelona, Spain

[&] These authors have contributed equally to this work

* romain.quidant@icfo.es; javier.garciadeabajo@icfo.es

**The much sought after optical transistor --the photonic counterpart of the electronic transistor-- is poised to become a central ingredient in the development of optical signal processing. The motivation for using photons rather than electrons not only comes from their faster dynamics but also from their lower crosstalk and minor environmental decoherence, which enable a high degree of integration and the realization of quantum operations** [1]. **A single-molecule transistor has been recently demonstrated at cryogenic temperatures** [2]. **Here, we demonstrate that a single NV centre at room temperature can operate as an optical switch under non-resonant CW illumination. We show optical modulation of more than 80% and time response faster than 100 ns in the green-laser-driven fluorescence signal, which we control through an independent near-infrared (NIR) gating laser. Our study indicates that the NIR laser triggers a fast-decay channel of the NV mediated by promotion of the excited state to a dark band. This simple concept opens a new approach towards the implementation of nanoscale optical switching devices.**

Unlike charged particles such as electrons, photons interact extremely weakly with each other [3]. Therefore, an optical switch requires the mediation of a physical system to produce efficient photon-photon interactions. Different approaches to optical transistors have been proposed [4][5][6], often based on nonlinearities in well-defined photonic resonators. More recently, a high-finesse optomechanical resonator has been used to demonstrate a transistor-like effect [7]. The ultimate goal for future optical commutation technologies is to rely on single atoms or molecules having the ability to manipulate light down to the single photon level, which feature an intrinsically high nonlinearity and grant us access to exploiting the quantum nature of light [8]. In this direction, a first single-molecule optical processing has been recently achieved at low temperatures [2] . Additionally, the inclusion of a third, dark state in the electronic structure of the atom or molecule has been argued to be particularly advantageous to reduce the effect of thermal decoherence [9][10]. The quest for operating such optical nano-transistor at room temperature is attracting considerable attention, as it would have major impact on photonic technologies and allow high-speed signal processing in general.

In parallel, research on single quantum emitters (either molecules or quantum dots (QDs)) has focused on achieving stable and efficient light emission with well-defined properties [11][12][13][14]. However a long-term stable source of single photons at room temperature still remains challenging with these emitters. In this context, Nitrogen Vacancies (NVs) in diamond have recently been intensely investigated due to their emission stability. NVs are artificial atoms protected from their environment by a diamond shell, and consequently, they are immune to both blinking and bleaching, so that their electron spin features a long coherence times, even at room temperature[15]. Their suitability for quantum-optics experiments has opened a new route towards the implementation of solid-state quantum simulators at room temperature [16] [17]. Despite the sustained effort to study the properties of NVs over the last decade [18], new insights on their energy states are still emerging [19][20]. Here, we describe an unconventional configuration in which a single NV operates as an all-optical nano-modulator at room temperature upon non-resonant continuous-wave (CW) near-infrared (NIR) illumination. Besides measuring and characterizing the switch effect, we provide insight into the mechanism responsible for the NV nonlinearity by revealing the presence of a dark band that can be optically activated with an IR laser. This new process is successfully described by a rate-equation model, further allowing us to predictively simulate and optimize the effect, which could be interesting for NV-centre-based optical quantum technologies.

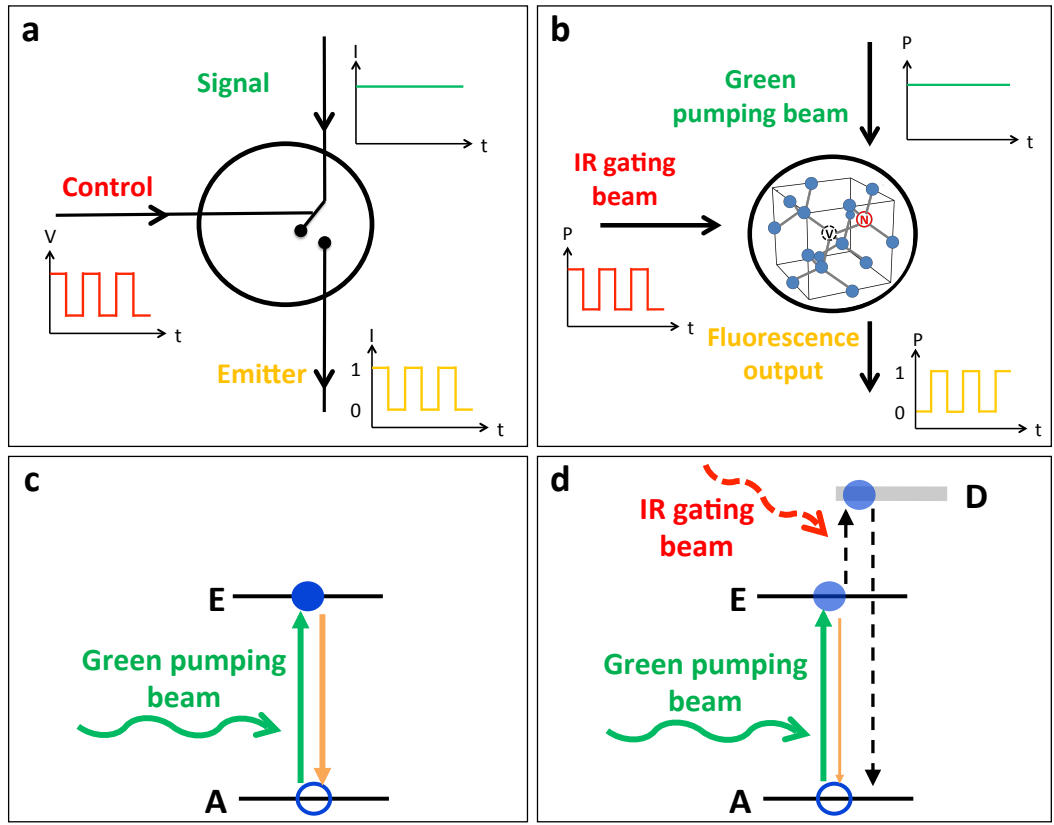

**Figure 1: Concept of optical switch based on a single NV centre.** **a** and **b,** Similar to its electronic counterpart (**a**), the emitter signal of the optical switch (NV-fluorescence upon continuous green-laser illumination) is controlled by a gate provided by a modulated non-resonant NIR laser (**b**). **c** and **d,** The NIR laser populates a dark band D, which decays non-radiatively, thus reducing the fluorescence output compared with the emission without NIR laser.

Our concept of optical modulator based on a single NV centre is sketched in Figure 1. The NV centre has a steady fluorescence (output signal) when pumped with a CW green laser (input signal). Illumination with a non-resonant CW NIR laser (gate signal) drives the excited state back to the ground state via non-radiative decay. In the following, we experimentally demonstrate the viability of this new gating channel, and speculate the potential involvement of a dark band of states to which the NIR laser promotes the excited state. This mechanism thus opens the possibility of controlling the number of emitted fluorescence photons by adjusting the intensity of the NIR gating laser.

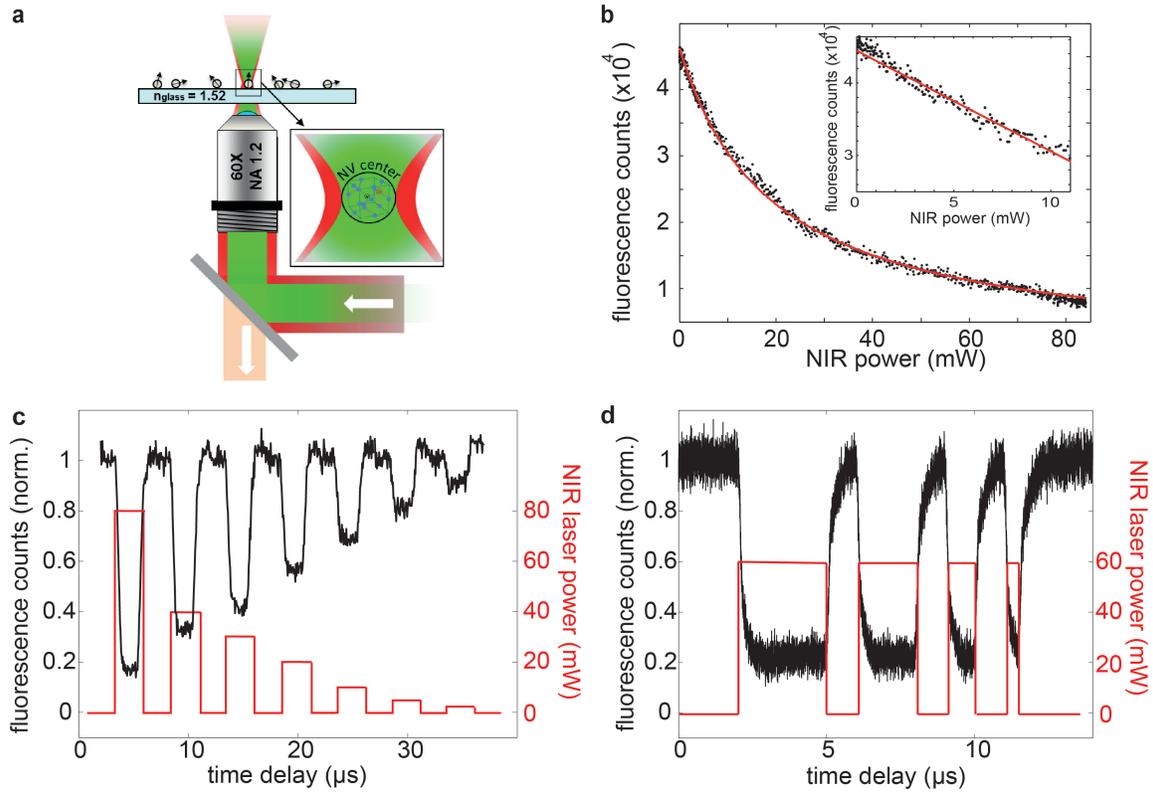

**Figure 2. Optical modulation of the NV luminescence under NIR illumination**. **a,** A green laser (λ=532 nm) is focused through a high numerical-aperture objective (NA=1.2) to excite a single NV centre in a nanodiamond crystal deposited on a glass coverslip (refractive index $n_{glass}$=1.52). A NIR laser (λ=1064 nm, shown in red in the sketch) is super-imposed to the 532 nm laser. The emitted fluorescence (orange) passes through a dichroic filter and is detected with two Avalanche PhotoDiodes (APDs). **b,** Setting the excitation power of the green laser to 55 µW while increasing the power of the NIR laser, a $1/(1+P_{NIR}/P_{sat})$ fluorescence decrease is observed, which is linear for small power (inset) and nonlinear when $P_{NIR}$ increases. **c** and **d,** By modulating the gating NIR laser with an Acousto-Optic Modulator (AOM) for constant green laser excitation power, a mimicking modulation of the NV fluorescence is observed. Tuning the power of the gating laser pulses allows us to reach different fluorescence decay levels (**c**). Here, we demonstrate the possibility of defining at least 8 states (i.e., well differentiated output levels) for potential application as a digital optical switch. The measured emission from the NV for different gating laser pulse lengths ranging from 3 µs down to 200 ns

illustrates the time response of the system (**d**). In our configuration, the response is faster than 200 ns and is limited by the electronic time response of our AOM.

In our experiment, 70nm nanodiamonds are dispersed on a glass coverslip that is placed in the sample plane of a scanning confocal fluorescence microscope. The NVs are pumped with a 532nm CW laser (input beam) after focusing through an immersion objective lens (NA=1.2), and their fluorescence (output beam) is collected back through the same objective and sent to an Avalanche PhotoDiode (APD) (Figure 2a). Handbury-Brown and Twist (HBT) measurements enable us to identify a nanodiamond that hosts a single NV (see Supplementary Information, SI). The 1064 nm laser beam (gating beam) is superimposed with the 532nm beam after passing through an Acousto-Optic Modulator (AOM) that allows us to temporally modulate its power (see Figure 2.c-d).

We start by examining the effect of 1064nm illumination on the fluorescence yield of a single NV centre under constant pumping with 55µW green light. The NIR power dependence of the fluorescence counts plotted in Figure 2b shows a monotonous drop that is apparently linear below 15mW (inset of Figure 2b) before reaching saturation. This demonstrates the possibility of modulating the fluorescence emission from a single NV by up to 80% for this chosen green excitation power. Quenching of fluorescence could be obtained with other molecular species by means of alternative mechanisms such as conformational transformation in photochromic molecules [21][22], although these are intrinsically much slower than the electronic response time of our system. Nevertheless, unlike molecules or quantum dots [11][13],

NV centers are stable and bright emitters, allowing us to observe this effect at room temperature and over a long time span [15][18]. A decrease in the fluorescence of NVs has been observed in prior studies upon Stimulated Emission Depletion (STED) under resonant excitation [23], multiphoton absorption induced by high power pulsed laser [24], local heating to hundreds of degrees [25], or even change of the nitrogen-vacnacy charge state [26]. In contrast, we here invoke a radically different physical mechanism, as emphasized by the fact that our non-resonant CW NIR illumination of a few tens of mW cannot trigger any of the above-mentioned mechanisms. Figure 2c shows the evolution of fluorescence from the same NV centre while applying a sequence of NIR laser pulses (3µs pulse width) with decreasing powers. The high stability of the NV emission enables us to define eight different output states. In this fashion, the NV centre gated by an NIR laser can be operated as a multi-level digital modulator. It is worth underlining that a single NV centre can be routinely used over several days without noticing any degradation in the emission properties. Also, identical results were repeated on several NVs from the same batch as well as on NVs from different providers.

In order to assess the time response of the switch effect, the NIR pulse duration was changed from 3 µs to 200 ns while maintaining a fixed NIR laser power (Figure 2d). The magnitude of the fluorescence drop is conserved for NIR pulses as short as 200 ns. We actually attribute this limit to the time response of the AOM used for the NIR laser modulation and expect the ultimate commutation rate to be only limited by the natural E→A decay rate of the NV.

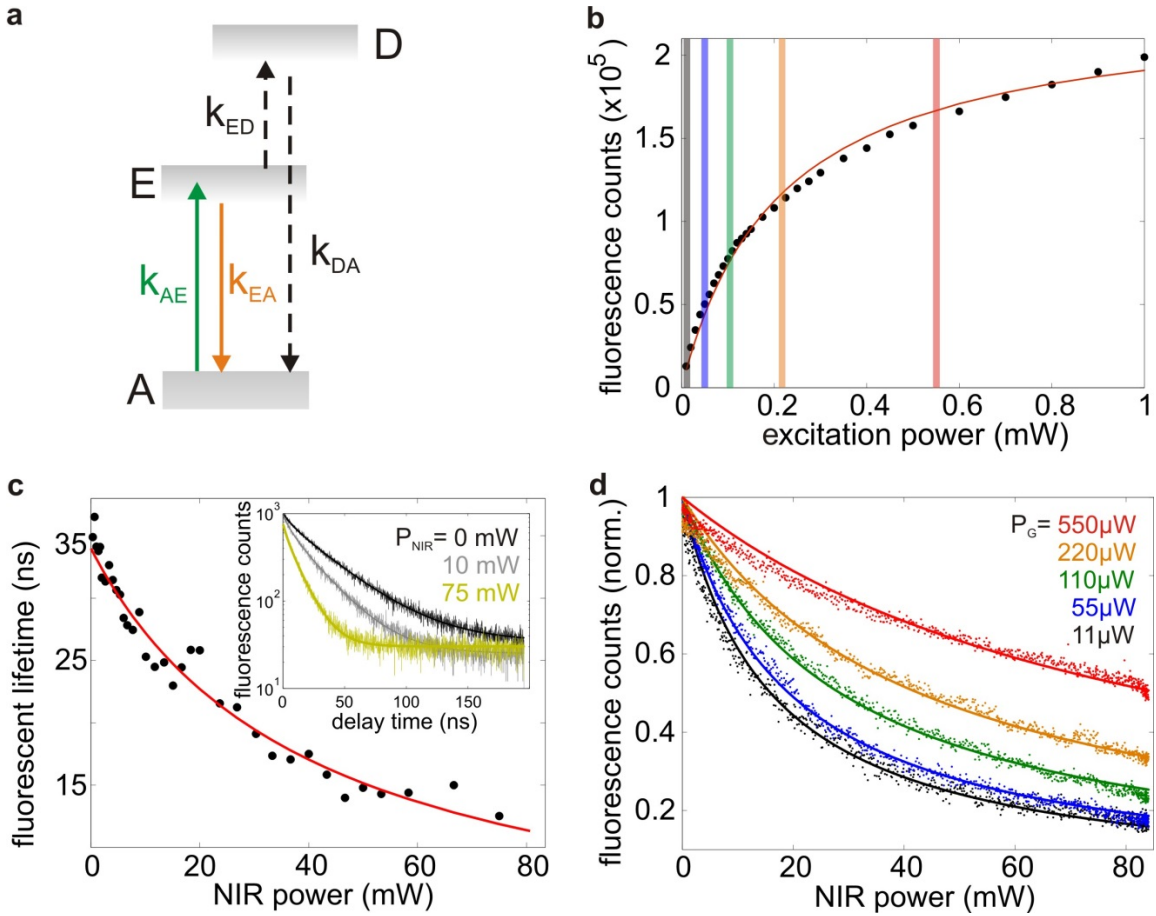

**Figure 3. Dependence of the fluorescence modulation of a single NV centre in diamond and its simulation. a,** Model of energy bands for an NV centre gated with an NIR laser, consisting of a triplet ground state A, a triplet excited state E and a fast decaying dark band D. **b,** Saturation of the NV fluorescence as a function of green-laser power $P_G$ in the absence of NIR laser ($P_{NIR}=0$). The red solid curve is a fit of the fluorescence $F$ to the formula extracted from the three-level rate equation model sketched in (**a**) (see SI): $F = \eta\, \kappa_{EA} \left( \frac{\alpha P_G}{1+\alpha P_G + \beta P_{NIR}} \right)$, where η is the detection efficiency, $\kappa_{EA}$ is the natural E→A decay rate and $\alpha$=0.009 µW$^{-1}$. **c,** Time-resolved measurements performed on a single NV centre after excitation with a 100 ps green laser under continuous NIR illumination. The measured power dependence of the excited-state lifetime (black dots) is extracted from the fits, as shown in the inset for three different NIR powers. The red solid curve is a fit of the fluorescence lifetime to the model, $\tau_E=(\kappa_{EA})^{-1}(1+\beta P_{NIR})^{-1}$, using β= 24.8 µW$^{-1}$. **d,** Normalised drop in background-corrected fluorescence with increasing NIR laser power for different levels of CW green-laser excitation power (see labels), corresponding to the vertical lines in (**b**) with the same colour code. The solid curves are a fit of the fluorescence signal *F* using the previously fitted values of parameters *α* and *β*. All data are acquired from the same single NV centre, and all theory curves are calculated for the same values of parameters *α* and *β*.

The dynamics and response of the fluorescence from a gated single NV is thoroughly analyzed in Figure 3, where we find excellent agreement with a model based upon a master rate equation involving just three energy bands (Figure 3a): the excited and ground states A and E, as well as a dark state D that is reached upon NIR illumination. After a few microseconds of CW green laser illumination, the system evolves towards the $m_s$=0 sublevel o the ground triplet state manifold [18], and thus, we can assume that states A and E are spin polarized, with a high probability into the $m_s$=0 levels of the ground and excited NV triplets. In fact, as most of the excitation events are associated with the intense and broad phonon sideband of the NV centre, narrow zero-phonon line features may be ignored in the fluorescence process. In what follows, we show that the NIR laser drives the system from E to A through an additional decay channel at very high rate compared with the natural decay rate $\kappa_{EA}$.

In the absence of NIR illumination, only states A and E are involved. Excitation of the system from A to E occurs at a rate $\kappa_{AE}=\alpha P_G \kappa_{EA}$ proportional to the green-laser power $P_G$, whereas decay from E to A proceeds via radiative decay with a rate $\kappa_{EA}$=1/(35 ns), that can change for different NV centers [27]. In our time-averaged measurements, it is safe to neglect quantum coherences and describe the system through a master rate equation, which leads to a fluorescence rate (see Supplementary Information, SI) $F= \eta \kappa_{EA}\ \alpha P_G /(1+\alpha P_G)$, where η = 0.005 is the detection efficiency of our system [28]. This produces an excellent fit to the green-laser-power-dependent fluorescence rate by choosing $\alpha=0.009$ $\mu W^{-1}$ (Figure 3b).

The effect of the NIR laser is best observed through the reduction that it produces in the lifetime of the excited state E (Figure 3c). The NV is first prepared in the $m_s=0$ manifold upon extended green laser illumination followed by a 1μs waiting time and is then excited with a 100 ps green-laser pulse. The ensuing time-resolved fluorescence reveals the lifetime, which is controlled by both natural decay at a rate $\kappa_{EA}$ and NIR-assisted decay at a rate $\beta P_{NIR}\kappa_{AE}$, must thus follow the expression $\tau_E=(\kappa_{EA})^{-1}(1+\beta P_{NIR})^{-1}$. This produces an excellent fit to the measured data with the choice $\beta= 24.8$ $\mu W^{-1}$ (Figure 3c).

The full dependence of the fluorescence rate on green-laser and NIR-laser powers predicted by the rate-equation model reduces to (see SI)

$$F = \eta\, \kappa_{EA} \left(\frac{\alpha P_G}{1+\alpha P_G+\beta P_{NIR}}\right). \qquad (1)$$

This simple expression is actually reproducing a detailed set of measured NIR-power-dependent profiles for different green-laser powers rather accurately with the single abovementioned choice of parameters $\alpha$ and $\beta$ (Figure 3d), thus corroborating the validity of our model. The main assumption of the model is the additional fast decay proportional to $P_{NIR}$, which is plausibly mediated by promotion to a dark band D. The nature of this band is an open question, although it is likely connected with higher-energy states reached via NIR photon absorption from the E band. Higher-energy states are more susceptible to decay non-radiatively, as they can access a larger number of levels through Auger and other electronic processes mediated by Coulomb interaction.

In our model, we assume a promotion rate $\kappa_{ED} = \beta P_{NIR}$ proportional to the NIR power, followed by fast decay from D to the ground state A. This is compatible with the sub-femtosecond time scale that is characteristic of non-radiative electronic transitions. Actually, the best fit to our data is obtained under the assumption $\kappa_{DA} \gg \kappa_{EA}$ (see SI).

The fast, efficient modulation of the emission of a single NV centre here presented demonstrates the principle of single-emitter optical switch operating at room temperature. A logical next step towards the exploitation of this concept for optical signal processing into an integrated scheme would consist in coupling a single NV to highly confined optical guided modes, as those supported for instance by a quasi-one-dimensional plasmonic waveguide[29][30].

**Acknowledgements**

This work was partially supported by the Spanish Ministry of Sciences (grants FIS2010–14834, MAT2010-14885 and CSD2007–046-NanoLight.es), the European Community's Seventh Framework Program under grant ERC-Plasmolight (no. 259196) and Fundació privada CELLEX. M. G. acknowledges the support of the FPU grant AP2009-3025 from the Spanish Ministry of Education. R. M. acknowledges support of Marie Curie and NEST programs. The authors thank M. Orrit for fruitful discussions.

**Author contributions**

M.G. and R.Q. conceived the experiment. M.G. and R.M. performed the experiments, J.G.A. provided theoretical support. All authors discussed the results and wrote the manuscript.

# Fast optical modulation of the fluorescence from a single NV centre

Michael Geiselmann [1, &], Renaud Marty [1, &], F. Javier García de Abajo [1,2], Romain Quidant [1, 2]

[1] ICFO - Institut de Ciencies Fotoniques, Mediterranean Technology Park, 08860 Castelldefels (Barcelona), Spain

[2] ICREA - Institució Catalana de Recerca i Estudis Avançats, Barcelona, Spain

[&] These authors have contributed equally to this work

**Supplementary information**

*S1: Fluorescence lifetime fit*

When dealing with nitrogen-vacancy (NV) centers located inside nanodiamonds (NDs), spin polarization is generally not perfect and both spin states coexist [1]. As the lifetime of the excited $m_s = \pm 1$ and $m_s = 0$ spin states are different, a bi-exponential fit of the lifetime is required to fit the experimental data [2][3]. The characteristic times of the two exponentials are associated with the fluorescent lifetime of each spin excited state and their amplitude coefficients give access to the spin polarization of the NV centre. In our case, we find that the spin polarization of the system is close to 80% and is not significantly modified with increasing NIR power (Figure S1). Even if spin polarization was not perfectly conserved under CW NIR illumination, the fluorescence drop observed cannot be attributed to a spin depolarization, since the fluorescence drop observed (up to 90%) is higher than the contrast between the $m_s = 0$ and $m_s = \pm 1$ states (up to 30%). As the system is mainly polarized into the spin state $m_s=0$, we then assume in our analysis that it can be modeled without taking into account the spin dependence of the NV centre in first approximation and Figure 3 of the main text represents the lifetime dependence of the $m_s=0$ sate.

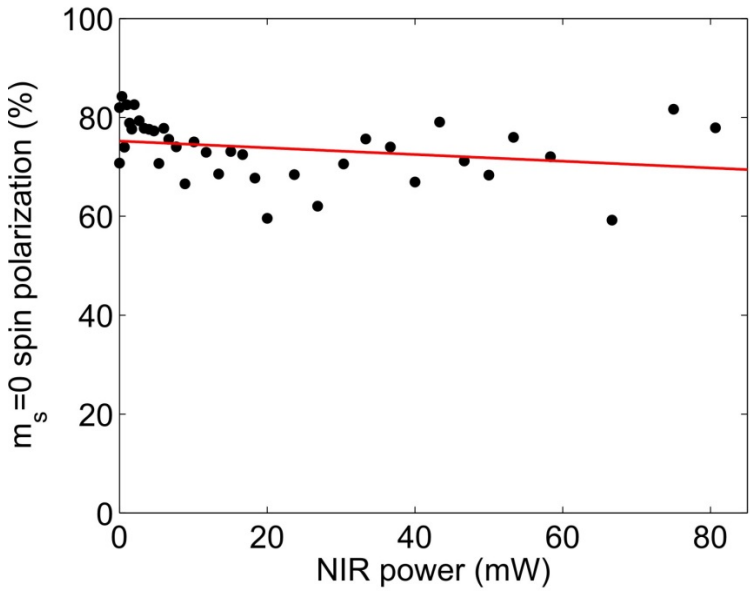

**Figure S1. Dependence of the spin polarization with the NIR power.** The values are extracted from the amplitude ratios of the bi-exponential fitted lifetime. The polarization is mainly kept in $m_s=0$ state whatever NIR power applied.

*S2: Second- order autocorrelation*

Figure S2 presents second-order autocorrelation measurements performed on the same single NV centre studied in the main article. The antibunching dip below 50 counts at zero delay time (Figure S2a, black line) demonstrates that it contains a single NV centre. For increasing excitation power (Figure S2) in the absence of NIR laser, bunching is observed when the power reaches the saturation level ($P_{sat}$= 110 µW, see Figure 3b). This effect, already observed in previous experiments on NV centers, is attributed to the population of a metastable state [4][5]. When exposing the NV centre to NIR illumination, we observe a reduction of the FWHM of the antibunching dip, which is consistent with the lifetime measurements displayed in Figure 3c. Additionally, the presence of the NIR laser leads to a small increase in bunching when the green excitation power is above saturation. This indicates that, beyond the linear excitation regime, the NIR laser could also lead to a modification of the intersystem crossing (ISC) rate toward the metastable state by saturating the NV centre. Nevertheless, the small effect observed on the bunching, even for high NIR power, indicates that the fluorescence modulation cannot be attributed to this effect. Moreover, the fast (below 100 ns) response of the effect here reported (see Figure 2d of main paper) and the excellent agreement obtained by assuming of a fast-decaying dark state in the model (see Sec. S3 below) both exclude an ISC mechanism.

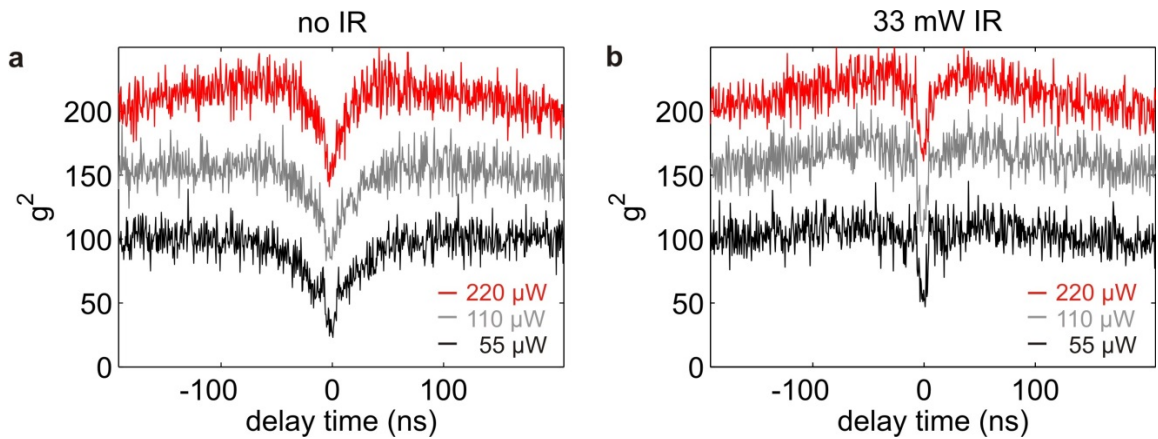

**Figure S2. Second-order autocorrelation of a single NV centre with and without NIR illumination. a,** Second-order autocorrelation taken without NIR illumination for different green excitation powers: below saturation (black), at saturation (grey) and above saturation (red). Bunching is observed for excitation powers above saturation. **b,** Second-order autocorrelation in the presence of 33 mW NIR illumination for the same green excitation power as in (a). The decrease in lifetime is confirmed by the reduction of the width of the antibunching dip. A small increase in bunching is observed, which we attribute to transfer of population to the metastable state.

*S3 NV centre gated by an NIR laser modeled as a three-band system*

We model the dynamics of the NV by means of a master rate equation in which we neglect quantum coherences. This is a reasonable approximation when discussing statistical measurements such as those reported here. We consider three energy levels (A, E and D) as shown in Figure 3a of the main paper. The population of these levels ($r_A$, $r_E$ and $r_D$) are assumed to be spin-averaged and we neglect effects arising from ISC. The dynamics of the system is then ruled by the master equation

$$\frac{d\rho_A}{dt} = -\kappa_{AE}\rho_A + \kappa_{EA}\rho_E + \kappa_{DA}\rho_D,$$

$$\frac{d\rho_E}{dt} = \kappa_{AE}\rho_A - \kappa_{EA}\rho_E - \kappa_{ED}\rho_E,$$

$$\frac{d\rho_D}{dt} = \kappa_{ED}\rho_E - \kappa_{DA}\rho_D,$$

where $\kappa_{EA}$ and $\kappa_{DA}$ are intrinsic decay rates, whereas $\kappa_{AE} = \alpha I_G \kappa_{EA}$ and $\kappa_{ED} = \beta I_{NIR} \kappa_{EA}$ are excitation rates assumed to be proportional to the green-laser and NIR-laser intensities, respectively, with proportionality constants $\alpha$ and $\beta$, expressed in units of the natural decay rate $\kappa_{EA}$. The steady-state solution of these equations is given by

$$\rho_A = \kappa_{DA}(\kappa_{EA} + \kappa_{ED}) / [\kappa_{DA}(\kappa_{EA} + \kappa_{ED}) + \kappa_{AE}(\kappa_{DA} + \kappa_{ED})],$$

$$\rho_E = \kappa_{AE}\kappa_{DA} / [\kappa_{DA}(\kappa_{EA} + \kappa_{ED}) + \kappa_{AE}(\kappa_{DA} + \kappa_{ED})],$$

$$\rho_D = \kappa_{AE}\kappa_{ED} / [\kappa_{DA}(\kappa_{EA} + \kappa_{ED}) + \kappa_{AE}(\kappa_{DA} + \kappa_{ED})].$$

From here, we obtain the fluorescence rate as

$$F = \eta \kappa_{EA} \rho_E = \eta \kappa_{EA} \frac{\alpha I_G}{1 + \alpha I_G + \beta' I_{NIR}},$$

where

$$\beta' = \beta \left(1 + \frac{\kappa_{EA}}{\kappa_{DA}} \alpha I_G \right)$$

and $\eta$ is the collection efficiency of the detector. Also, the lifetime of E is obtained from the rate equations by studying the exponential decay of $r_E$ when the system is prepared as $r_E = 1$ and $r_A = r_D = 0$ at $t = 0$. This leads to the intuitive result

$$\tau_E = \kappa_{EA}^{-1} \frac{1}{1 + \beta I_{NIR}}.$$

When fitting our experimental data in Figure 3 of the main paper, we find that the best fit is obtained under the assumption of a large ratio $\kappa_{DA}/\kappa_{EA} \gg 1$, so that we find an $I_G$-independent parameter $\beta' = \beta$. The experimental data are thus consistent with a NIR-assisted fast decay channel (much faster than the radiative decay from E to A), as expected from non-radiative electronic decay from D to A mediated by Coulomb interaction.

*S4 Working point of the modulator:*

As a maximum drop of fluorescence counts is observed for low green excitation powers, a tradeoff in this system is to find a power that results in enough fluorescent counts and at the same time has the largest drop in fluorescence.

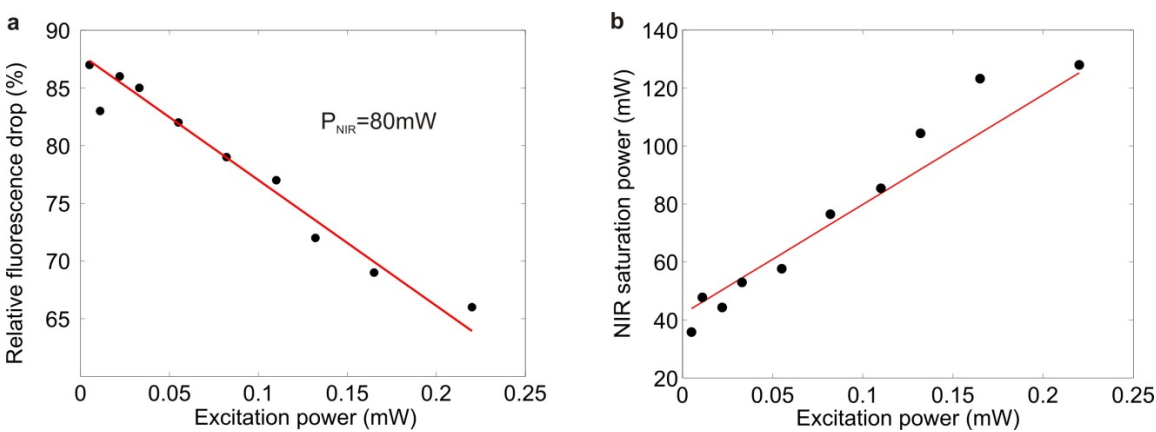

**Figure S3. a,** Fluorescence drop with increasing green excitation power. **b,** NIR power needed for a 70 % drop of fluorescence as a function of green excitation power.

Figure S3a shows the evolution of the normalized fluorescence drop induced by a NIR laser as a function of the green excitation power for fixed NIR gating power $P_{NIR}$=80 mW (see Figure 3d). The drop reduces when increasing the green excitation power. Additionally, Figure S3b shows that the NIR power required to reach a 70 % fluorescence drop. The required NIR power increases linearly when the green excitation power is increased. That means that the lower the excitation green power, the smaller the gating NIR power required to achieve large fluorescence modulation. Nevertheless, it is important to notice that a reduction of the green laser power leads to a decrease in fluorescence counts (as shown on the saturation curve in figure 3b). This implies that the power requirement by such modulator is limited by the sensitivity of the fluorescence detection and the coupling efficiency.

*S5 Temperature dependence upon NIR illumination*

We here report Optically Detected Magnetic Resonance (ODMR) spectroscopy [1][2][7] measurements that allow us to rule out any temperature-related effect as the reason for the fluorescence drop. This technique relies on the possibility of optically reading out the spin state of a NV center by tracking the fluorescence emission while sweeping the frequency of an applied microwave field over the spin transition frequencies. When the microwave frequency is resonant with a spin transition, a drop in fluorescence is observed. In nano-diamonds, two resonances are usually observed because internal strains break the degeneracy between the $m_s=0$ and $m_s=\pm1$ transitions. Several groups have recently shown a temperature dependence of the electronic spin in NV centers [3], [8–11]. It was found that the zero field splitting (ZFS) parameter D of the spin Hamiltonian is shifted by 75 kHz/K [8], as extracted from an electron spin resonance (ESR) measurement. An additional experiment shows that the ESR contrast of the resonances decreases with increasing temperature [3]. Here we perform ESR measurements to show that our decrease in fluorescence is not related to a substantial increase in temperature.

Fig. S4 a) and b) display the ESR measurements performed on two different individual NV centers as a function of NIR laser irradiance. The presence of the NIR laser induces two main features. First, a clear decrease of the ESR contrast is observed in Fig S4 c) and d) for increasing NIR laser power. This is in good agreement with the experiments reported in the main paper, where we show that the presence of the NIR laser produces a drop in fluorescence yield for the $m_s=0$ spin state and consequently reduces the difference between the $m_s=\pm 1$ and $m_s=0$ fluorescence intensities. A similar ESR contrast decrease has already been observed for temperatures above 600 K [3]. In order to measure the temperature of the NV center, we have then analyzed the shift of the ZFS parameter D. Fig S4 e) and f) show the dependence of D on NIR power. We find a linear relation of 17 kHz/mW for NV1 and 6.3 kHz/mW for NV2, which imply a temperature increase of 7 K and 3 K, respectively, at 30 mW.

This result confirms that the fluorescence switching is not related to a temperature increase. In fact, the temperature increases by less than 10 K for a 30 mW NIR power, for which we already have a fluorescence decrease of 60 % (Fig. 2c). This amount of heating is typical of experiments involving focusing a mW NIR beam to a diffraction-limited spot.

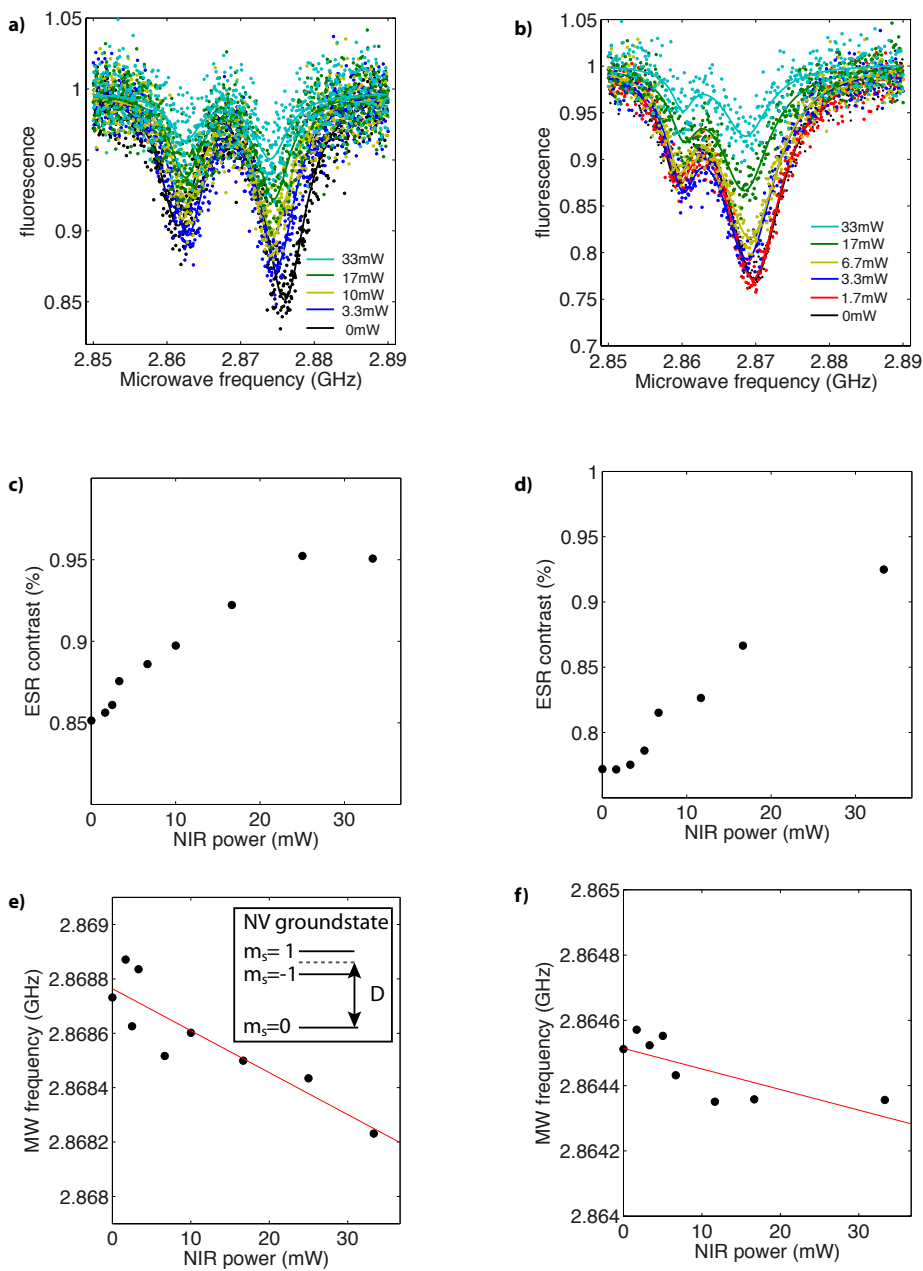

**Figure S4. NIR power dependence of ESR for individual NV centers. a,b,** ESR measurement for two different NV centers. Each graph shows a microwave frequency scan across the ground state transition frequencies for $m_S= 0$ and $m_S= \pm 1$. No magnetic field is applied. The splitting is due to strain inside the nanodiamond. **c,d,** ESR contrast in dependence of the NIR power. For increasing NIR power, the ESR contrast decreases as the difference in fluorescence intensity between the two spin states is decreasing due to the dark state excited by the NIR illumination. **e,f,** Dependence of the zero field splitting parameter D (see inset) on NIR power showing a thermally induced frequency shift of 75kHz/K [8], which yields less than 10 K temperature increase for 30 mW illumination power.